\documentclass[aps,pra,twocolumn,superscriptaddress,showpacs,floatfix]{revtex4-1}
\usepackage{amsmath}
\usepackage{graphicx}
\usepackage{dcolumn}
\usepackage{bm}
\usepackage{amssymb}
\usepackage{mathrsfs}
\usepackage{subfigure}
\usepackage{braket}
\usepackage[colorlinks,
            linkcolor=blue,
            anchorcolor=blue,
            citecolor=blue,
            urlcolor=blue
            ]{hyperref}
\usepackage{xcolor}	 
\definecolor{myred}{RGB}{255,66,56}

\definecolor{myblue}{RGB}{34,31,217}

\newcommand{\be}{\begin{equation}} 
\newcommand{\ee}{\end{equation}}

\newenvironment{mymathfrac}[2]{\raise0.5ex\hbox{$#1$}\! \left/ \! \lower0.5ex\hbox{$#2$}\right.}

\begin{document}

\title{Geometrical meaning of winding number and its characterization of topological phases in one-dimensional chiral non-Hermitian systems}
\date{\today}
\author{Chuanhao Yin}
\affiliation{Beijing National Laboratory for Condensed
Matter Physics, Institute of Physics, Chinese Academy of Sciences, Beijing 100190,
China}
\affiliation{School of Physical Sciences, University of Chinese Academy of Sciences, Beijing, 100049, China}

\author{Hui Jiang}
\affiliation{Beijing National Laboratory for Condensed
Matter Physics, Institute of Physics, Chinese Academy of Sciences, Beijing 100190,
China}
\affiliation{School of Physical Sciences, University of Chinese Academy of Sciences, Beijing, 100049, China}

\author{Linhu Li}
\affiliation{Department of Physics, National University of Singapore, 117542, Singapore}

\author{Rong L\"{u}}
\affiliation{Department of Physics, Tsinghua University, Beijing 100084, China}
\affiliation{Collaborative Innovation Center of Quantum Matter, Beijing, China}

\author{Shu Chen}
\thanks{Corresponding author: schen@iphy.ac.cn}
\affiliation{Beijing National Laboratory for Condensed
Matter Physics, Institute of Physics, Chinese Academy of Sciences, Beijing 100190,
China}
\affiliation{School of Physical Sciences, University of Chinese Academy of Sciences, Beijing, 100049, China}
\affiliation{Collaborative Innovation Center of Quantum Matter, Beijing, China}
\date{ \today}

\begin{abstract}

We unveil the geometrical meaning of winding number and utilize it to characterize the topological phases in one-dimensional chiral non-Hermitian systems. While chiral symmetry ensures the winding number of Hermitian systems being integers, it can take half integers for non-Hermitian systems. We give a geometrical interpretation of the half integers by demonstrating that the winding number $\nu$ of a non-Hermitian system is equal to half of the summation of two winding numbers $\nu_1$ and $\nu_2$ associated with two exceptional points respectively. The winding numbers $\nu_1$ and $\nu_2$ represent the times  of real part of the Hamiltonian in momentum space encircling the exceptional points and can only take integers. We further find that the difference of $\nu_1$ and $\nu_2$ is related to the second winding number or energy vorticity. By applying our scheme to a non-Hermitian Su-Schrieffer-Heeger model and an extended version of it, we show that the topologically different phases can be well characterized by winding numbers. Furthermore, we demonstrate that the existence of left and right zero-mode edge states  is closely related to the winding number $\nu_1$ and $\nu_2$.
\end{abstract}

\maketitle

\section{Introduction}
Recently there has been a growing interest in the study of topological properties of non-Hermitian Hamiltonian systems \cite{2007-Bender-p947,Esaki2011,Liang2013,2015-Malzard-p200402,Lee2016,Leykam2017,Menke2017,Xu2017,Molina2017, Guo2018,Hu2017,Xiong2017,Runder2009, Zhu2014, Zhou2017,Lieu,Alvarez,Zeng2016,Song2016}.
In comparison with Hermitian systems, non-Hermitian systems exhibit a special spectral degeneracies known as exceptional points (EPs) \cite{2001-Heiss-p149,Rotter2009,Heiss2012,2004-Berry-p1039,2004-Dembowski-p056216,Hassan,Hu2017}, where the Hamiltonian becomes non-diagonalizable, making it have complex band structure \cite{Shen2017} and achieve different properties from Hermitian systems. In Hermitian systems, Hermiticity grants the real eigenvalue and the eigenvector orthogonality, while in non-Hermitian systems, eigenvalue can be complex and Hamiltonian admits a complete biorthonormal system of eigenvectors when it is diagonalizable\cite{Mostafazadeh-JPA}. A number of theoretical works have indicated that various non-Hermitian models may support topologically nontrivial properties. For example, non-Hermitian analogs of Su-Schrieffer-Heeger (SSH) model either with parity-time ($\mathcal{PT}$) symmetry \cite{Zhu2014,Menke2017,2015-Yuce-p1213,Hu2011} or without PT symmetry have been studied, and recent experimental realizations of the PT-symmetry SSH model have demonstrated the existence of robust edge states. More recently, particular attention has been paid to a one-dimensional (1D) non-Hermitian model with its Hamiltonian in the momentum space encircling an EP \cite{Lee2016}, which stimulates the intensive theoretical studies of the characterization of topological phases and bulk-boundary correspondence in non-Hermitain systems \cite{Lee2016,Leykam2017,Menke2017,Xiong2017,Lieu,Alvarez}.

For 1D topological systems, the winding number can be used to characterize the topological properties of the $\mathcal{Z}$-class insulators, which is closely related to the quantized Berry phase (Zak phase) \cite{Zak1989} and protected by chiral symmetry \cite{Li2015}. In Hermitian systems, the winding number $\nu$ is always an integer. While $\nu=0$ describes the topologically trivial state, $\nu=\pm$1 correspond to topological states with zero-energy edge states.
When the momentum runs over the Brillouin zone (BZ), the topologically nontrivial system is characterized by a quantized Berry phase $\pi$ with a modulus of $2 \pi$, and correspondingly the Hamiltonian in the momentum space encircles the original point. The number of times encircling the original point is described by the winding number. It has been demonstrated that some extended topological models with long-range hopping terms support topological phases characterized by winding number with $\nu>1$ and exhibit rich phase diagram  \cite{Niu2012,Song2015,Li2015}. In contrast to the Hermitian topological system, the geometric meaning of winding number of non-Hermitian systems is still not very clear. Moreover, although there exist debates on the validity of the bulk-boundary correspondence, it may still be an important principle in topological non-Hermitian systems \cite{Esaki2011, Leykam2017, Lee2016}. \par

In this work, we study the geometrical meaning of the winding number and explore the relation between the winding number and zero-mode edge states by studying a non-Hermitian SSH model and its extension with more rich topological phases described by higher winding numbers. The winding number of non-Hermitian Hamiltonian $\nu$ is found to be equal to half of the summation of two winding numbers $\nu_1$ and $\nu_2$, which describe the times of trajectory of the real part of the non-Hermitian Hamiltonian surrounding two EPs, respectively. From the trajectory picture, it is clear that the winding numbers  $\nu_1$ and $\nu_2$ can only take integers. We also find that the difference of $\nu_1$ and $\nu_2$ is equal to twice of the second winding number or energy vorticity defined in previous works \cite{Leykam2017,Shen2017}. By studying the non-Hermitian SSH and extended SSH models and scrutinizing the zero-mode solutions of semi-infinite systems under open boundary condition (OBC), we further demonstrate that $\nu_1$ and $\nu_2$ correspond to zero-mode edge states at different boundaries, i.e.,  $\nu_1$ gives the number of zero-mode states at the left edge and $\nu_2$ at the right edge. \par

The paper is organized as follows. In Sec.\ref{sec:winding}, we introduce the definition of winding number surrounding EPs and derive its relation with the winding number of the non-Hermitian Hamiltonian. In Sec. \ref{sec:ssh} and Sec.\ref{sec:extend-ssh}, two examples (non-Hermitian SSH model and extended non-Hermitian SSH model) are explored in details. Topologically different phases in the phase diagram are determined via the calculation of winding numbers and the relation between the winding number and the number of zero-mode edge states is also discussed. A brief summary is given in Sec.\ref{sec:summary}.

\section{Winding number for chiral non-Hermitian system} \label{sec:winding}
We consider a general two-band non-Hermitian system, whose Hamiltonian in momentum space only contains two of the three Pauli matrices and can be written in the form of
\begin{equation}
h(k) = h_x \sigma_x + h_y \sigma_y
\end{equation}
after some rotations, with $\sigma$ the Pauli matrices. Here $k$ is the momentum, and $h_x$ and $h_y$ are generally functions of $k$. This model has chiral symmetry as
\[
\sigma_z h(k) \sigma_z = -h(k).
\]
When $k$ varies from $0$ to $2\pi$, $h_x$ and $h_y$ form a close loop. Thus we can generalize the definition of the winding number for the Hamiltonian in the parameter space spanned by $h_x$ and $h_y$:
\begin{eqnarray}
\nu &=& \frac{1}{2\pi}\oint_c \frac{h_x dh_y-h_y dh_x}{h_x^2+h_y^2} \\
    &=&\frac{1}{2\pi}\int_{0}^{2\pi}dk \frac{h_x \partial_k h_y-h_y \partial_k h_x}{h_x^2+h_y^2}, \label{nu}
\end{eqnarray}
where $c$ is a close loop with $k$ varying from $0$ to $2\pi$.
The winding number for this chiral-symmetry-protected system can be associated to the non-Hermitian Zak phase
\be
\gamma=\int_{0}^{2\pi}dk \frac{\langle u_k^L|i\partial_k|u_k^R\rangle}{\langle u_k^L| u_k^R\rangle},
\label{zak}
\ee
via the relation $\gamma= \nu \pi$, where $u_k^{L(R)}$ denote the occupied left and right Bloch state eigenvectors of the Hamiltonian. The eigenstate for the band with eigenvalue $-\sqrt{h_x^2+h_y^2}$ then has the form of $\langle u_{k}^L|=\frac{1}{\sqrt{2}}(\frac{h_x+ih_y}{\sqrt{h_x^2+h_y^2}},-1)$ and $|u_{k}^R \rangle=\frac{1}{\sqrt{2}}(\frac{h_x-ih_y}{\sqrt{h_x^2+h_y^2}},-1)^T$.

For the non-Hermitian system, $h_x$ and $h_y$ are general complex or at least one of them is complex, so we introduce a complex angle $\phi$ satisfying $\tan \phi = h_y/h_x $. In terms of $\phi$, the winding number $\nu$ can be represented as
\be
\nu = \frac{1}{2\pi}\oint_c \partial_k\phi dk, \label{eq:winding_angle}
\ee
where the integral is also taken along a loop with $k$ from $0$ to $2\pi$. Eq.(\ref{nu}) is always well defined, except at the EPs, where we have $h_x^2 + h_y^2 = 0$, giving rise to the location of EPs at
\begin{equation}
h_{xr} = -h_{yi}, ~~and~~h_{yr} = h_{xi} \label{condition1}
\end{equation}
or
\begin{equation}
h_{xr} = h_{yi}, ~~and~~
h_{yr}= - h_{xi},  \label{condition2}
\end{equation}
where $h_{xr} = Re(h_x)$, $h_{xi} = Im(h_x) $, $h_{yr} = Re(h_y) $ and $h_{yi} = Im(h_y) $. It means that in this system, the number of EPs is no more than two. 
When the system is Hermitian with $h_x$ and $h_y$ being real,  these two EPs merge into the original point $(h_x, h_y) = (0,0)$, which is the gap closing point in the Hermitian spectrum.
Except at EPs, $\phi$ is well defined as a function of $k$, and we always have
\be
e^{i2\phi}=\frac{1+i\tan\phi}{1-i\tan\phi}=\frac{h_x+ih_y}{h_x-ih_y}.
\ee
Since $\phi$ is a complex angle, we can decompose it into two parts
\be
\phi =\phi_r +i\phi_i, 
\ee
where $\phi_r = Re(\phi)$ and $\phi_i=Im(\phi)$. Thus, Eq.(\ref{eq:winding_angle}) becomes the integral of two parts. In $e^{i2\phi }=e^{i2\phi_r}e^{-2\phi_i}$, $\phi_r$ contributes to argument, while $\phi_i$ contributes to amplitude,
\begin{eqnarray*}
e^{-2\phi_i } &=&\left\vert \frac{h_{x}+ih_{y}}{h_{x}-ih_{y} }\right\vert. \\
\end{eqnarray*}
As $\phi_i $ is a real continuous periodic function of $k$, we have
\begin{eqnarray*}
\oint_c \partial_k\phi_i dk = \phi_i(2\pi) - \phi_i(0)=0,
\end{eqnarray*}%
which means the imaginary part of $\phi$ having no effect on the integral of winding number. On the other hand,
\begin{displaymath}
e^{i2\phi_r }
=\mymathfrac{\displaystyle \frac{h_{x}+ih_{y}}{h_{x}-ih_{y} }}{\displaystyle \left\vert \frac{h_{x}+ih_{y}}{h_{x}-ih_{y} }\right\vert }.
\end{displaymath}
By using the relation
\begin{eqnarray*}
\tan 2\phi_r  &=&\mymathfrac{\displaystyle Im \frac{h_{x}+ih_{y}}{h_{x}-ih_{y} }}{\displaystyle Re \frac{h_{x}+ih_{y}}{h_{x}-ih_{y} } }, \\
\end{eqnarray*}
after some algebras, we can rewrite the above relation as
\begin{eqnarray}
\tan 2\phi_r  &=&\frac{\tan \phi _{1}+\tan \phi _{2}}{1-\tan \phi _{1}\tan
\phi _{2}} = \tan \left( \phi _{1}+\phi _{2}\right) . \label{totalangle}
\end{eqnarray}
where
\begin{eqnarray} \label{tan_phi}
\tan \phi _{1} =\frac{h_{yr}+h_{xi}}{h_{xr}-h_{yi}}, ~~\tan \phi _{2} =\frac{h_{yr}-h_{xi}}{h_{xr}+h_{yi}},
\end{eqnarray}%
which define two real angles $\phi_1$ and $\phi_2$, respectively.

From Eq.(\ref{totalangle}), it is straightforward to get the relation between $\phi_r$ and $\phi_1$, $\phi_2$, i.e., $\phi_r =(\phi _{1}+\phi _{2})/{2} + n\pi$, where $n$ is an integer. Substituting them into Eq.(\ref{eq:winding_angle}) and using the relation $\oint_c \partial_k\phi_i dk =0$, we get
\begin{equation}
\nu =\frac{1}{2} (\nu_1 +\nu_2 ), \label{nutot}
\end{equation}
where
\begin{eqnarray}
\nu_1 &=& \frac{1}{2\pi}\oint \partial_k\phi _{1}dk   \label{nu1} \\
\nu_2 &=& \frac{1}{2\pi}\oint \partial_k\phi _{2}dk  \label{nu2}
\end{eqnarray}
can be viewed as two winding numbers. In the space spanned by $h_{xr}$ and $h_{yr}$, the point $(h_{yi},-h_{xi})$ corresponds to one of the EPs, and $\phi_1$ is the angle of $(h_{xr}, h_{yr})$ relative to this EP. This indicates that $\nu_1$ is the winding number of the real part of Hamiltonian surrounding the EP point $(h_{yi},-h_{xi})$ when $k$ ranges from $0$ to $2\pi$. Similarly, $\nu_2$ represents the winding number about the other EP point $(-h_{yi},h_{xi})$.

\begin{figure}[tb]
\centering
\includegraphics[width=0.95\columnwidth]{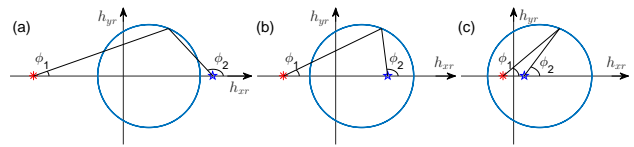}
\caption{\label{fig:example} (Color online) A Schematic diagram of three typical winding cases by calculating the model with $h_x = t+t'\cos(k)$ and $h_y = t'\sin(k) - i\delta$. In (a)-(c) $t=0.5$, $t'=1$ and $\delta$ takes $1.75$, $1$  and $0.2$, respectively. Times for the trajectory surrounding the EP marked by the asterisk gives $\nu_1$ and the one surrounding the pentagram gives $\nu_2$. It is straightforward to get (a) $\nu_1=0$ and $\nu_2=0$, (b) $\nu_1=0$ and $\nu_2=1$ and (c) $\nu_1=1$ and $\nu_2=1$. }
\end{figure}

To get an intuitive understanding, in Fig.\ref{fig:example}, we show three typical winding cases for a simple model with $h_x=t + t' \cos k$ and $h_y=t' \sin k -i \delta$, which corresponds to the non-Hermitian SSH model to be considered in next section. In Fig.\ref{fig:example}(a), these two EPs are not surrounded by the trajectory of Hamiltonian and its winding number is 0. While Fig.\ref{fig:example}(b) and (c) are corresponding to winding number $\nu=1/2$ and $1$, as one EP and two EPs are surrounded in Fig.\ref{fig:example}(b) and (c), respectively.
The winding number $\nu$ describes half the total number summation of times that the real part of Hamiltonian travels counterclockwise around two EPs. Despite that the non-Hermitian Zak phase only takes $0$, $\pi/2$ or $\pi$ with a modulus of $2\pi$, the winding number can be any half-integer, and indicates different topological phases. A more complicate example with $\nu=3/2$ and $2$ shall be studied by considering an extended non-Hermitian SSH model in section \ref{sec:extend-ssh}.

Before moving on to the study of concrete models, we would like to demonstrate that our introduced winding number $\nu_1$ and $\nu_2$ also have a correspondence to the second winding number $w_2$ defined in Ref.\cite{Leykam2017}, and the energy vorticity defined in Ref. \cite{Shen2017}. Consider a quantity $\nu'$ defined as
\begin{equation} \label{nu'}
\nu' \equiv \frac{\nu_1-\nu_2}{2},
\end{equation}
we obtain
\begin{eqnarray*}
\nu' =  \frac{1}{2\pi}\oint \partial_k \frac{ \phi _{1}- \phi _{2}}{2} dk = \frac{1}{2\pi} \oint \partial_k\phi' dk,
\end{eqnarray*}
where $2\phi'=\phi_1-\phi_2$.Then we have
\begin{eqnarray*}
\tan 2\phi' = \tan (\phi_1-\phi_2)
= \frac{\tan \phi_1 - \tan \phi_2}{1+\tan \phi_1 \tan \phi_2}.
\end{eqnarray*}
Substituting Eq.(\ref{tan_phi}) into it, we obtain
 \begin{eqnarray*}
\tan 2\phi' = \frac{Im ( h_x^2+h_y^2 )}{Re ( h_x^2+h_y^2 ) }
= \tan Arg (h_x^2+h_y^2),
\end{eqnarray*}
which indicates that $\phi' = \frac{1}{2} Arg(h_x^2+h_y^2) + n\pi$ with $n\in \mathbb{Z}$. Thus the new quantity $\nu'$ takes the form as
\begin{eqnarray}
\nu' &=& \frac{1}{2\pi} \oint \partial_k \Big[ \frac{1}{2} Arg(h_x^2+h_y^2) + n\pi \Big]dk \nonumber\\
&=& \frac{1}{2\pi} \oint \partial_k Arg \sqrt{h_x^2+h_y^2} dk \label{nu'-2},
\end{eqnarray}
where $\sqrt{h_x^2+h_y^2}$ is one of the two eigen-energies of the two-band system. This definition of $\nu'$ is equivalent to the definition of the second winding number $w_2$ in Ref.\cite{Leykam2017}.
From Eqs.(\ref{nutot}) and (\ref{nu'}), we can also get
\begin{equation}
\nu_1 = \nu + \nu', ~~ \nu_2 = \nu -\nu'.
\end{equation}

\section{Non-Hermitian SSH model} \label{sec:ssh}
\begin{figure}[tb]
\centering
\includegraphics[width=0.9\columnwidth]{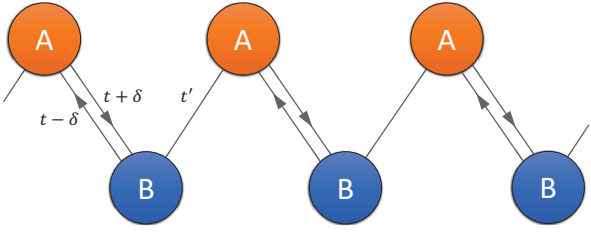}
\caption{\label{fig:chain} (Color online) A Schematic diagram of non-Hermitian SSH model with different nearest-neighbor hopping strength inside a cell.}
\end{figure}

To give a concrete example and demonstrate our scheme, first we consider the non-Hermitian version of SSH model \cite{Lieu,SSH,Linhu2014} by changing the hopping term in the unit cell into non-Hermitian form with different strengths in the right and left hopping directions as shown in Fig.\ref{fig:chain}, described by the Hamiltonian
\begin{eqnarray}
H &=& \sum_{n}\left[(t-\delta)\hat{a}^{\dagger}_{n}\hat{b}_{n}  + (t+\delta)\hat{b}_n^\dagger \hat{a}_n \right. \nonumber \\
& & \left. ~~~ +t'\hat{a}^{\dagger}_{n+1}\hat{b}_{n}+t'\hat{b}_{n}^\dagger \hat{a}_{n+1} \right].
\end{eqnarray}
where $\hat{a}_n^\dagger$, $\hat{b}_n^\dagger$ ($\hat{a}_n$, $\hat{b}_n$) are the creation (annihilation) operators at the $n$-th $A$, $B$ site. After Fourier transform, we have
\be
H = \sum_k \psi^\dagger_k h(k) \psi_k,
\ee
where $\psi_k=(a_k , b_k)^T $ and
\begin{equation}
h(k) = \left(\begin{array}{cc}
0 & t' e^{-ik}+t-\delta \\
t' e^{ik}+t+\delta & 0
\end{array}\right). \nonumber
\end{equation}
For simplicity, we take $t'=1$, $t$ and $\delta$ real. In general, when $\delta = 0$, this model reduces to the standard SSH model, which belongs to $Z-$type topological system with topological and trivial phase characterized by the winding number $\nu=1$ and $0$, respectively. The dispersion of this Hamiltonian is
\begin{eqnarray*}
E = \pm \sqrt{1+t^2-\delta^2+2t\cos k-i2\delta \sin k}.
\end{eqnarray*}
From the expression of dispersion, we see that the energy is symmetric about zero energy, which is ensured by the chiral symmetry. Since the energy gap must close at phase transition points, we can determine the phase boundaries of the non-Hermitian SSH model by the band crossing condition $E(k)=0$, which yields $t = \pm \delta + 1$ and $t=\pm \delta - 1$.

\begin{figure}[tbp]
\centering
\includegraphics[width=0.9\columnwidth]{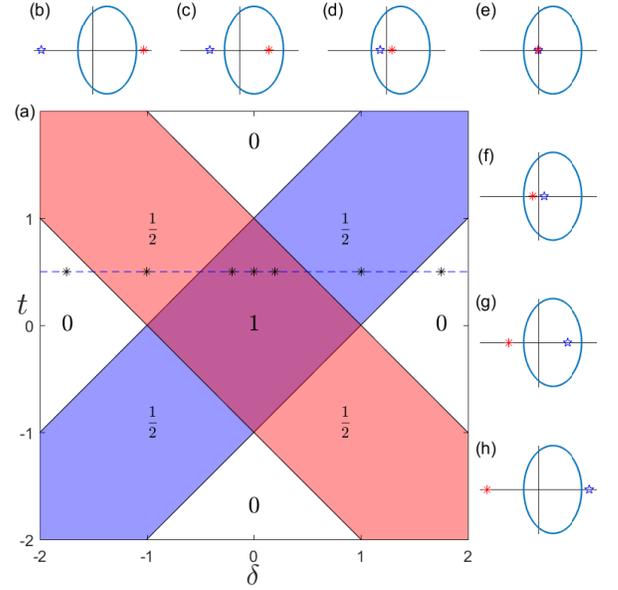}
\caption{\label{fig:winding-ssh} (Color online) (a) The phase diagram of the non-Hermitian SSH model. The colors from shallow to deep represent $\nu=0$, $1/2$ and $1$ as marked. (b)-(h) show winding diagrams with different parameters. They are all on the dash line in (a) with $t=0.5$  and different values of $\delta$. In each subfigure, (b) $\delta=-1.75$, (c) $\delta=-1$, (d) $\delta=-0.2$, (e) $\delta=0$, (f) $\delta=0.2$, (g) $\delta=1$ and (h)$\delta=1.75$. Red asterisk and blue pentagram represent two different EPs.}
\end{figure}

By using Eqs.(\ref{nu}) and (\ref{nutot}), we calculate the winding number of this model, as shown in Fig.\ref{fig:winding-ssh}(a). Topologically different phases in the phase diagram can be distinguished by their winding numbers $\nu=0$, $1/2$ and $1$. In Fig.\ref{fig:winding-ssh}(b)-(h), we illustrate the winding of the projection of Hamiltonian with different $\delta$ by fixing $t=0.5$. While the red asterisk represents the EP of $(-\delta,0)$, the blue pentagram represent the other EP of $(\delta,0)$. When the momentum $k$ varies from $0$ to $2\pi$, the curve of the Hamiltonian may enclose both of two EPs for Figs.(d)-(f) with $\nu_1=\nu_2=1$, or only one of the EP with $\nu_1=1$ for Fig.(c) or $\nu_2=1$ for Fig.(g), or not enclose them with $\nu_1=\nu_2=0$ for Fig.(b) and Fig.(h), which correspond to $\nu=1$ ,$1/2$ and $0$, respectively. However, at the phase transition points, the curve crosses any of the EPs, and the winding number is ill defined due to $h_x^2+h_y^2 = 0 $ at these points. Except these phase transition points, the total enclosed times of EPs is twice of the winding number $\nu$.

Next, we discuss existence of zero modes under OBC.
In real space, we consider a semi-infinite system with $n=1,2,3,...$, and choose the wavefunction as $\Psi_{m}=\sum_{n}(\psi_{a,n}\hat{a}_n^{\dagger}+\psi_{b,n}\hat{b}_n^{\dagger})|0\rangle$, with $|0\rangle$ the vacuum state. From $H\Psi_{m}=E_{m}\Psi_{m}$, we have the eigenequations:
\begin{eqnarray}
(t+\delta) \psi_{a,n}+\psi_{a,n+1}=E_m\psi_{b,n} \label{eigenssh_b},\\
(t-\delta) \psi_{b,n}+\psi_{b,n-1}=E_m\psi_{a,n} \label{eigenssh_a}.
\end{eqnarray}
Here, for $n=1$, we have $\psi_{b,0}=0$, which gives the left boundary condition.
\begin{figure}[tb]
\centering
\includegraphics[width=0.95\columnwidth]{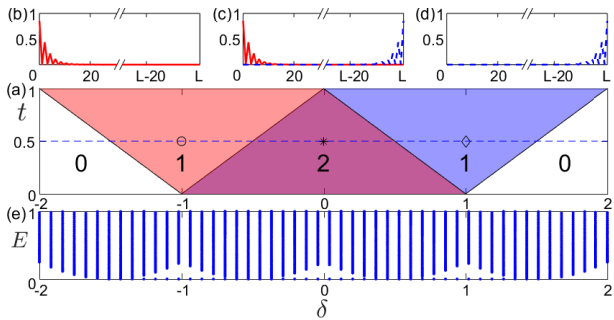}
\caption{\label{fig:edge-ssh} (Color online)
(a) Phase diagram of non-Hermitian SSH model characterized by the number of zero-mode edge states. The red shallow represents the existence of left zero-mode edge state, while the blue shallow represents the right edge state. The number represents the quantities of edge state in its mark area. Subgraphs (b)-(d) are distributions of moduli of the zero-mode edge states of the three marks (ring, star and square) at $t=0.5$ and $\delta=-1,0,1$ in subgraph (a), respectively.
(e) The dispersion of $H^\dagger\times H$ with parameters $t=0.5$ and $\delta$ changing from $-2$ to $2$ for the lattice size with 200 sites.
 }
\end{figure}

A zero-mode solution of the system corresponds to $E_m=0$. When $E_m=0$,  Eqs.(\ref{eigenssh_b}) and (\ref{eigenssh_a}) are decoupled. Now Eq.(\ref{eigenssh_a}) becomes
\begin{eqnarray}
(t-\delta) \psi_{b,n}+\psi_{b,n-1}=0, \label{eigenEq_d}
\end{eqnarray}
and we have $\psi_{b,n}=0$ for any $n$ by using the above relation and the boundary condition $\psi_{b,0}=0$. Similarly, Eq.\ref{eigenssh_b} can also be simplified as
\begin{eqnarray}
(t+\delta)\psi_{a,n}+\psi_{a,n+1}=0, \label{eigenEq_c}
\end{eqnarray}
which gives an exponentially decreasing state only if $|t+\delta|<1$, i.e., there exists a solution localized at the left edge when $|t+\delta|<1$. We note that here we choose a semi-infinite geometry with only one edge. When the system has a finite size with two ends, edge states with non-zero $b$ component localized at the other end can exist when $E_m=0$ and $|t-\delta|<1$. \par

In Fig.\ref{fig:edge-ssh}(a), we show the phase diagram characterized by the number of zero-mode edge states, which is consistent with the phase diagram determined by the winding number. Different color in the shadow area represents different distribution of the edge state. In the red area with $|t+\delta|<1$, there exists only one zero-mode edge state which is localized at the left side as shown in Fig.\ref{fig:edge-ssh}(b) for the system with $t=-1$ and $\delta = 0.5$.  In the blue area with $|t-\delta|<1$, there exists an edge state localized at the right side as shown in Fig.\ref{fig:edge-ssh}(d) for system with $t=1$ and $\delta = 0.5$. In the middle crossing area, there exist two zero-mode edge states, which appear at both sides as shown in Fig.\ref{fig:edge-ssh}(c) corresponding to $t=0$ and $\delta = 0.5$. It is also straightforward to see the relation between winding number and the zero-mode edge states. While $\nu=1/2$ corresponds to the existence of one zero-mode edge state, $\nu=1$ corresponds to two zero-mode edge states. Although both $(\nu_1=1, \nu_2=0)$ and $(\nu_1=0, \nu_2=1)$ correspond to the same $\nu=1/2$, they give rise to different $\nu'$ with $\nu'=1/2$ and $\nu'=-1/2$ respectively. Our results unveil that winding on different EP relates to different edge state, i.e., $\nu_1=1$ or $\nu_2=1$ corresponds to the zero-mode edge state at the left or right edge, respectively. In the region of $\nu=0$, we do not find the existence of zero-mode edge solution.

Alternatively, we can check whether the zero-mode solution exists by diagonalizing the Hermitian operator $H^\dagger\times H$. If  $H$ has zero mode solutions, then $H^\dagger$ also has zero modes. Thus, we could use $H^\dagger\times H$ to illustrate the existence of zero mode in $H$. We note that when $H$ holds zero mode, $H^\dagger\times H$ must hold zero mode, but on the contrary it may not be true. The spectrum of $H^\dagger\times H$ corresponding to the blue dash line in Fig.\ref{fig:edge-ssh}(a) is shown in Fig.\ref{fig:edge-ssh}(e). It is shown that the zero-mode solution obtained by solving Eq.(\ref{eigenEq_d}) and (\ref{eigenEq_c}) is consistent with the zero mode in the spectrum shown in Fig.\ref{fig:edge-ssh}(e). \par

\section{Extended non-Hermitian SSH model} \label{sec:extend-ssh}
Next, we consider an extended non-Hermitian SSH model, which takes the form
\be
H = \sum_k \psi^\dagger_k h(k) \psi_k,
\ee
where $\psi_k=(a_k, b_k)^T $ and
\begin{equation}
h(k) = \left(\begin{array}{cc}
0 & t-\delta+t'e^{-ik}+\Delta e^{-2ik}\\
t+\delta+t' e^{ik}+\Delta e^{2ik} & 0
\end{array}\right). \nonumber
\end{equation}
For simplicity, we take $t'=1$, and $t$, $\delta$, $\Delta$ real. In general, when $\Delta = 0$, this model reduces to non-Hermitian SSH model. The dispersion of this Hamiltonian is
\begin{eqnarray*}
E = \pm \sqrt{(t-\delta+e^{-ik}+\Delta e^{-2ik})(t+\delta+e^{ik}+\Delta e^{2ik})}.
\end{eqnarray*}
From the expression of dispersion, we see that the energy is symmetric about zero energy, which is ensured by the chiral symmetry. Since the energy gap must close at phase transition points, we can determine the phase boundaries of the non-Hermitian SSH model by the band crossing condition $E(k)=0$, which yields $t =\pm \delta + 1 -\Delta$ and $t =\pm \delta - 1 -\Delta$ for arbitrary $\Delta$ and additionally $t=\Delta \pm \delta$ if $|\Delta|>0.5$. \par
\begin{figure}[tbp]
\centering
\includegraphics[width=0.8\columnwidth]{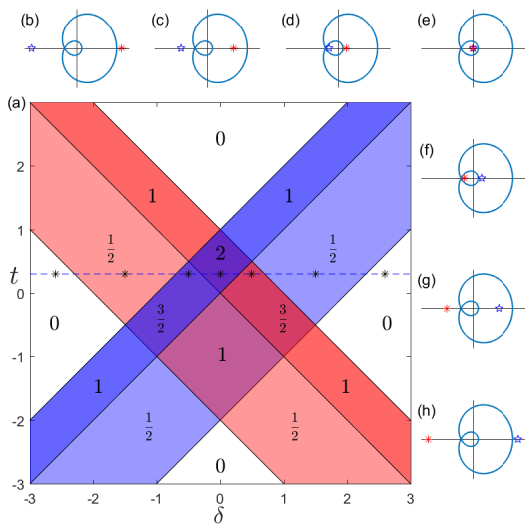}
\caption{\label{fig:winding-exssh} (Color online)(a) The phase diagram of the non-Hermitian extended SSH model with $\Delta=1$. The colors from shallow to deep represent $\nu=0$, $1/2$, $1$, $3/2$ and $2$ as marked. (b)-(h) show winding diagrams with different parameters. They are all on the dash line in (a) with $t=0.3$  and different values of $\delta$. In each subfigure, (b) $\delta=-2.6$, (c) $\delta=-1.5$, (d) $\delta=-0.5$, (e) $\delta=0$, (f) $\delta=0.5$, (g) $\delta=1.5$ and (h) $\delta=2.6$. Red asterisk and blue pentagram represent two different EPs.}
\end{figure}
In Fig.\ref{fig:winding-exssh}(a), we show the winding number calculated by Eq.(\ref{nu}) with $\Delta=1$.
Topologically different phases in the phase diagram are distinguished by their winding numbers $\nu=0$, $1/2$, $1$, $3/2$ and $2$.
In Fig.\ref{fig:winding-exssh}(b)-(h), we also illustrate the winding trace of the projection of Hamiltonian with different $\delta$ by fixing $t=0.3$, corresponding to $\nu=0, 1/2, 3/2, 2, 3/2, 1/2, 0$, respectively.
From the pattern of winding trace, we can easily read out the information of winding numbers $\nu_1$ and $\nu_2$ and thus $\nu = (\nu_1 + \nu_2)/2$. For example, we have $\nu_1=1$ and $\nu_2=0$ for Fig.\ref{fig:winding-exssh}(c) and $\nu_1=1$ and $\nu_2=2$ for  Fig.\ref{fig:winding-exssh}(d), corresponding to $\nu =1/2$ and $3/2$, respectively.
\par 

Next we explore the zero-mode solution of the extended non-Hermitian SSH model under OBC. In real space, the Hamiltonian reads
\begin{eqnarray}
H=\sum_{n}\Big[ &&(t-\delta)\hat{a}^{\dagger}_{n}\hat{b}_{n}+(t+\delta) \hat{b}_n^\dagger \hat{a}_n +  \hat{a}^{\dagger}_{n+1}\hat{b}_{n} \nonumber \\
&&+ \hat{b}_{n}^\dagger \hat{a}_{n+1} + \Delta \hat{a}_{n+2}^\dagger \hat{b}_n + \Delta \hat{b}_n^\dagger\hat{a}_{n+2} \Big].
\end{eqnarray}
Here we also consider a semi-infinite system with $n=1,2,3,...$, and choose the wavefunction as $\Psi_{m}=\sum_{n}(\psi_{a,n}\hat{a}_n^{\dagger}+\psi_{b,n}\hat{b}_n^{\dagger})|0\rangle$. $H\Psi_{m}=E_{m}\Psi_{m}$ yields
\begin{eqnarray}
(t+\delta) \psi_{a,n}+\psi_{a,n+1}+\Delta \psi_{a,n+2}=E_m\psi_{b,n} \label{eigenexssh_be},\\
(t-\delta) \psi_{b,n}+\psi_{b,n-1}+\Delta \psi_{b,n-2}=E_m\psi_{a,n} \label{eigenexssh_ae}.
\end{eqnarray}
Since $n$ starts from 1, here we set $\psi_{b,n} = 0$ for $n\leqslant 0$ as the left boundary condition.

\begin{figure}[tbp]
 \centering
\includegraphics[width=0.9\columnwidth]{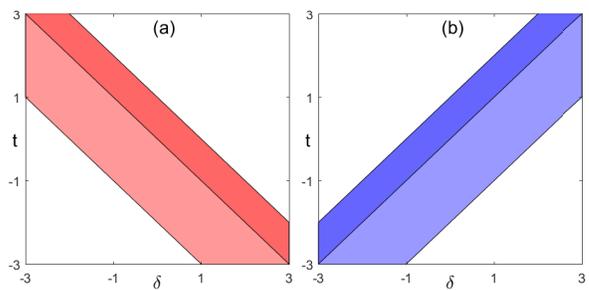}
\caption{\label{fig:edge-location} (Color online) (a) In this figure, we distinguish different regions by the number of
left zero-mode edge states for the semi-infinite system with the end at left side. In the dark area, there are two left zero-mode edge states corresponding to the region with both eigenvalues less than unity, while in the shallow area, there is only one zero-mode edge state. In the blank area, there is no zero-mode state. Subfigure (b) shows the case with the end at the right side. Similarly, blank, shallow and dark areas corresponds to regions with no, one and two right zero-mode edge states.}
\end{figure}

A zero-mode solution can be achieved by setting $E_n = 0$, as
\begin{eqnarray}
(t+\delta)\psi_{a,n}+\psi_{a,n+1}+\Delta\psi_{a,n+2}=0,\\
(t-\delta)\psi_{b,n}+\psi_{b,n-1}+\Delta\psi_{b,n-2}=0.
\end{eqnarray}
The boundary condition indicates $\psi_{b,n} = 0$ for every $n$. For edge states localized at $A$ sublattice, we expect it to decay exponentially, thus we assume $\psi_{a,n+1}/\psi_{a,n}=\lambda$, which leads to
\begin{eqnarray} \label{lam}
\Delta\lambda^2+\lambda+(t+\delta)=0.
\end{eqnarray}
The quadratic equation gives two complex solutions $\lambda_1$ and $\lambda_2$, and we can obtain an edge state localized at the left end of the 1D semi-infinite chain whenever one of modules of $\lambda_i$ ($i=1,2$) is less than unity.
If there are two eigenvalues less than unity, there must be two edge states at this side. If only one eigenvalue is less than unity, there exists one edge state. If no eigenvalue is less than unity, no zero-mode edge state exists at the left side.
A general solution of Eq.(\ref{lam}) is given by $\lambda = \frac{-1\pm\sqrt{1-4\Delta(t+\delta)}}{2\Delta}$. For $\Delta=1$,
it is easy to find that when $-\delta<t<1-\delta$, both solutions of $\lambda$ are less than unity, indicating the existence of two left zero-mode edge states corresponding to $\nu_1=2$. When $-2-\delta<t<-\delta$, only one of these two solutions is less than unity, indicating the existence of one left zero-mode edge state corresponding to $\nu_1=1$. In other cases, no $\lambda$ is less than unity. To see it clearly, we show the regions with one solution and two solutions of $\lambda$ being less than unity corresponding to the shallow and dark red areas in Fig.\ref{fig:edge-location}(a), with one and two edge states located at the left side, respectively.
Similar method can also be applied to obtain the condition for the existence of one and two right zero-mode edge states by solving a semi-infinite system with the end at right side, which yields
\begin{eqnarray} \label{lam2}
\Delta\lambda'^2+\lambda'+(t-\delta)=0.
\end{eqnarray}
with the right edge at the $L-$th $B$ site, where $\lambda' = \psi_{b,n-1}/\psi_{b,n}$ and $\psi_{a,n}=0$.
Similarly, we get a diagram as shown in Fig.\ref{fig:edge-location}(b), in which the shallow and dark blue areas correspond to regimes with one and two edge states located at the right side, respectively. Combining Fig.\ref{fig:edge-location}(a) and (b) together, we get a phase diagram characterized by the number of left and right zero-mode edge states, which
is consistent with the phase diagram shown in Fig.\ref{fig:winding-exssh}(a).

\begin{figure}[tbp]
 \centering
\includegraphics[width=0.8\columnwidth]{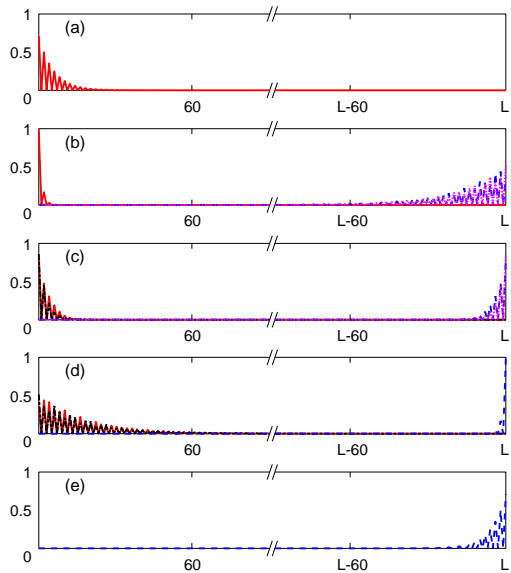}
\caption{\label{fig:edge-exssh} (Color online) (a)-(e) are distribution of moduli of the zero-mode edge states wavefunction under open boundary condition, which correspond to the situation in subgraph (c)-(g) in Fig.\ref{fig:winding-exssh} with $\Delta =1$, $t=0.3$ and $\delta = -1.5,-0.5,0,0.5,1.5$ respectively. Red solid lines and black dash-dot lines represent left-side edge states, while blue dashed lines and purple dotted lines represent right-side edge states. }
\end{figure}

In Fig.\ref{fig:edge-exssh}, we show the distribution of moduli of zero-mode wavefunctions for systems corresponding to Fig. \ref{fig:winding-exssh}(c)-(g) from top to bottom, respectively, e.g., Fig. \ref{fig:edge-exssh}(a) shows that there is only one edge state localized at the left boundary, corresponding to Fig.\ref{fig:winding-exssh}(c) with $\nu=1/2$ and $\nu_1=1$, and  there are two edge states localized at the right boundary and one at the left boundary in Fig. \ref{fig:edge-exssh}(b), corresponding to Fig.\ref{fig:winding-exssh}(d) with $\nu=3/2$, $\nu_1=1$ and $\nu_2=2$, {\it etc}.   All the solutions are consistent with the phase diagram shown in Fig.\ref{fig:winding-exssh}.\par

\section{Summary} \label{sec:summary}
In summary, we have explored the geometrical meaning of winding number for general 1D two-band non-Hermitian systems with chiral symmetry and applied it to determine the phase diagram of a non-Hermitian SSH model and an extended version of it. By generalizing the definition of winding number for Hermitian systems to non-Hermitian systems, we unveil that the introduced winding number $\nu$ is equal to half of the summation of two winding numbers $\nu_1$ and $\nu_2$, i.e., $\nu =(\nu_1+\nu_2)/2$, where  $\nu_1$ and $\nu_2$ can only take integers as they represent the winding times for the close trajectory of the real part of non-Hermitian Hamiltonian surrounding two EPs when the momentum goes across the BZ. We further demonstrate that the difference of $\nu_1$ and $\nu_2$ is equal to the twice of the second winding number or energy vorticity $\nu'$ introduced in Ref.\cite{Leykam2017} and Ref.\cite{Shen2017}, i.e.,  $\nu' =(\nu_1-\nu_2)/2$.
We then determine the phase diagrams of the non-Hermitian SSH and extended SSH models by calculating the winding numbers. It is shown that different topological phases can be well characterized by the winding numbers. By studying the zero-mode solutions of semi-infinite systems for these two non-Hermitian models, we find that the existence of left and right zero-mode edge states is closely related to the winding numbers $\nu_1$ and $\nu_2$.
While $\nu_1+\nu_2$ indicates the total number of the zero-mode edge states, values of $\nu_1$ and $\nu_2$  provide the information of where these edge states localize.
Our work gives a geometrical interpretation for the winding numbers of 1D non-Hermitian systems and establishes connection between the number of zero-mode edge states and the winding numbers. Our scheme can be applied to other 1D  non-Hermitian models with chiral symmetry. \par

Note added: After the submission of our work, a work appeared on the arXiv \cite{Gong2018}, in which topological classification of non-Hermitian systems according to Altland-Zienbauer classes is studied. While our work is focused on the 1D non-Hermitian systems with chiral symmetry, the definitions of winding numbers $\nu$, $\nu_1$ and $\nu_2$ can not be directly generalized to systems without chiral symmetry.

\begin{acknowledgments}

The work is supported by NSFC under Grants No. 11425419, the National Key Research and Development Program of China (2016YFA0300600 and 2016YFA0302104) and the Strategic Priority Research Program (B) of the Chinese Academy of Sciences  (No. XDB07020000). We thank C. Yang for helpful discussions.
\end{acknowledgments}

\end{document}